\documentstyle[epsf]{l-aa}
\begin{document}
\newcommand{\bold}[1]{\mbox{\protect\boldmath$#1$}}
\newcommand{\rK}{{\mbox{\tiny K}}}
\newcommand{\rd}{{\rm d}}
\newcommand{\beq}{\begin{equation}}
\newcommand{\eeq}{\end{equation}}

\thesaurus{02.01.2; 05.03.1}
\title{An orbiter crossing an accretion disc}
\author{Ladislav \v{S}ubr \and
        Vladim\'{\i}r Karas}
\offprints{L.~\v{S}ubr}
\institute{Astronomical Institute, Charles University Prague, 
V~Hole\v{s}ovi\v{c}k\'ach~2, CZ-180\,00~Praha, Czech~Republic\\
E-mail: subr@kocour.ms.mff.cuni.cz; vladimir.karas@mff.cuni.cz}

\date{Received ..............  ...., 1999; accepted .................}

\maketitle

\markboth{L.\ \v{S}ubr \& V.\ Karas: An orbiter crossing an accretion
 disc}{L.\ \v{S}ubr \& V.\ Karas: An orbiter crossing an accretion disc}

\begin{abstract}
We further investigate the long-term evolution of a trajectory of a
stellar-mass orbiter which is gravitationally bound to a massive central
body acting by Newtonian force. The orbiter undergoes repetitionary
collisions with an accretion disc. We consider eccentric orbits
intersecting the disc once or twice per each revolution and we solve the
equations for osculating elements. We find the terminal radii of the
orbits and time needed to bring the orbit into the disc plane as
function of initial parameters, and we show that previous simplified
estimates (derived for the case of low eccentricity) remain valid within
factor of two. The discussion presented in this paper offers a toy model
of the orbital evolution of a satellite which passes through the
rarefied gaseous environment of a stationary disc. We demonstrate that
the drag on the satellite should be taken into account in calculations
of the stellar distribution near the super-massive black hole in
galactic centers. On the other hand, it does not impose serious
restrictions on the results of gravitational wave experiments, except in
the case of discs with rather high surface density.


\end{abstract}

\section{Introduction}
Rapid stellar motion in galactic centers reflects properties of a
presumed supermassive compact body in the nucleus (Binney \& Tremaine
1987; Merrit \& Quinlan 1998a, b). One can define the sphere of
influence, $r_{\rm{}h}=GM/\langle\sigma\rangle^2$, which determines the
region of space where gravity of the centre dominates over more remote
parts of the galaxy ($M\ga10^6M_{\odot}$ is the central mass,
$\sigma\approx10^2\,$km/s is velocity dispersion). Inside $r_{\rm{}h}$,
a star can be considered as a satellite orbiting around the central
body. We introduce gravitational radius $r_{\rm{}g}=2GM/c^2$, which
characterizes compactness of that body and provides a natural
length-scale for the problem discussed in this paper. When expressed in
terms of $r_{\rm{}h}$ and $r_{\rm{}g}$, we will explore in this paper
the region $r_{\rm{}g}\ll{r}\ll{r_{\rm{}h}}$.

\begin{figure}[t]
 \epsfxsize=\hsize
 \centering
 \mbox{\epsfbox{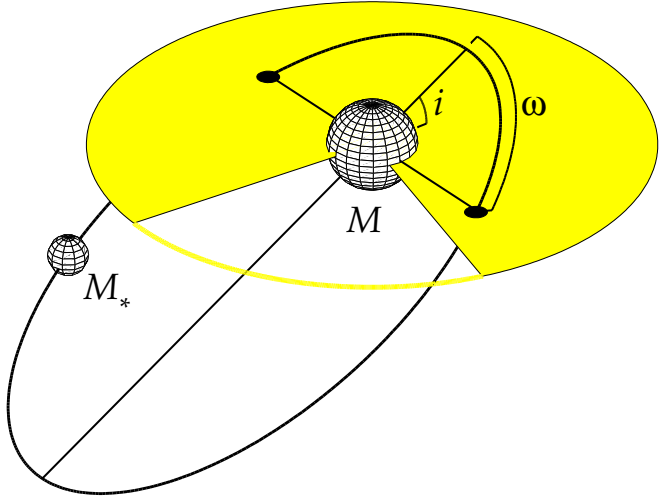}}
 \caption{Geometry of the model. The satellite (mass
 $M_{\ast}{\approx}M_\odot$) follows a slowly evolving elliptical orbit
 around more massive central body ($M{\approx}10^8M_\odot$). Argument of
 pericentre $\omega$ and inclination angle $i$ specify orientation of
 the ellipse. In the present model, action of the disc matter upon the
 satellite depends only on the disc properties at the places of
 intersections. For $\omega=\pi/2$ (as in this sketch) both
 intersections with the disc occur at the same distance, while the case
 is just opposite for $\omega=0$ (the places of intersections then
 correspond to the pericentre and the apocentre, respectively).
 \label{fig0}}
\end{figure}

Several mechanisms act as perturbations on almost free motion of
orbiters in the central gravitational field inside $r_{\rm{}h}$: gravity
of the background galaxy, dynamical friction in the field of other
orbiters, energy losses due to gravitational radiation, tidal
interactions, and friction caused by the rarefied gaseous environment
(see Rees 1998 for a recent review). Here we will concentrate on the
latter effect, namely, we will consider the long-term evolution of
orbital parameters (eccentricity $e$, semimajor axis $a$, inclination
$i$, argument of pericentre $\omega$) of the orbiter which crosses the
plane of an accretion disc (Fig.~\ref{fig0}). We assume that the disc is
geometrically thin (collisions will be treated within the impulsive
approximation) and neglect all other dissipative effects, ablation of
the orbiter, corrections due to general relativity, as well as
self-gravity of the disc. Some of these effects were examined quite
recently by Vokrouhlick\'y \& Karas (1998b), Ivanov et al.\ (1998), and
Zurek et al.\ (1994). In this context, gradual evolution of the disc
surface density was explored by Ivanov et al.\ (1999). Further citations
to older works can be found in Syer et al.\ (1991), Vokrouhlick\'y \&
Karas (1993), and Artymowicz (1994).

The impulsive approximation ignores effects of resonances, and it
assumes nonzero inclination of the orbit with respect to the disc plane.
Our calculations thus do not apply to the case when the orbiter remains
embedded inside the disc, the situation which has been widely discussed
because of its profound implications for describing the formation of
stars and planets (Lin et al.\ 1999; Bryden et al.\ 1999). Also, we do
not consider possible disruption of the orbiter by the central body;
this process has been discussed by many authors, e.g.\ Kim et al.\
(1999) very recently. Under these assumptions, the orbital evolution can
be solved in two steps: First, parametric relations for osculating
elements are obtained which determine the shape of the orbit ($e(i),$
$a(i)$, etc). Next, the temporal dependences are examined. Although the
present discussion is very much simplified, it captures basic properties
of the mechanism driving the orbital evolution, and it offers a toy
model which can be used as a test bed for more complicated and
astrophysically realistic studies.

In this paper we extend discussion of Vokrouhlick\'y \& Karas (1998a;
cited as VK hereafter) who studied the orbital evolution under the
assumption that orbiter's trajectory intersects the disc only once per
revolution. Such an assumption applies to the case when the disc has a
finite radial extent and the orbiter follows an eccentric path (the
initial stage of the capture from an energetically unbound state).
Regarding the description of the disc matter and its mutual interaction
with the orbiter, VK adopted several assumptions which are quite
plausible for this type of the toy model: a geometrically thin disc
rotating at Keplerian velocity in purely azimuthal direction. It was
further assumed that the disc itself does not change with time; in
particular, dissipative waves induced by the orbiter and the possibility
of opening of a gap was not not taken into account. These simplifying
assumptions are maintained also in the present paper.

Gradual circularization and precession lead eventually to the orbit with
two intersections per revolution; at this moment the analytical solution
from VK ceases to be valid. Terminal stages of the evolution were treated
by Rauch (1995) under approximation of zero and moderate eccentricity,
but orbits with large eccentricity could not be fully examined in terms
of an analytic solution, because the two intersections can occur at
different radial distances. One thus needs to specify surface density of
the disc as a function of radius. We write down corresponding
evolutionary equations and solve them numerically under the assumption
of the power-law surface density profile. We examine the temporal
evolution of orbital parameters, showing that a very special orientation
of the trajectory, $\omega=\pi/2$ (when both intersections are at the
same $r$), represents a stable configuration which can be treated
analytically. In other words, our present discussion extends previous
discussion of the initial period of the orbital evolution, and it
concerns also the intermediate stage, when two intersections per
revolution have already developed while inclination is still nonzero.

\section{Equations for orbital parameters}

The basic formulation of the problem follows VK, in particular, the
osculating elements of the orbiter are updated at every intersection
with an axially symmetric disc, which is described by surface density
$\Sigma_\rd\,\propto\,r^s$ ($s={\rm{const}}$) and by orbital velocity
$\bold{v}_\rd=v_{\phi}\,\bold{e_\phi}$. It appears natural to consider
Keplerian rotation of the disc, $v_\phi\,\equiv\,v_\rK=\sqrt{GM/r}$,
which fits with the assumption about its small geometrical thickness,
and enables us to introduce vertically integrated quantities in the
standard manner (we examined also different rotation laws; Karas \&
\v{S}ubr 1999). The central body attracts the orbiter by Newtonian
force. Passages through the disc are treated as inelastic collisions
which, by assumption, do not affect the properties of the disc and of
the orbiter itself. Stated in a different way, disc material located
along the orbiter's path transfers part of its momentum to the orbiter,
but then the disc restores its original stationary state before the next
collision occurs. Collisions thus lead to tiny changes of the orbiter's
velocity, $\bold{v}\rightarrow\bold{v'}$. The momentum conservation law
gives
\beq
 \bold{v'}=\frac{1}{A+1}\Big[v_r\,\bold{e_r}
 +v_{\vartheta}\,\bold{e_\vartheta}
 +\left(v_\varphi+Av_\rK\right)\bold{e_\phi}\Big], \label{eq:v'}
\eeq
where
$A(r)\,\equiv\,{\Sigma_\rd}v_{\rm{rel}}\Sigma_\ast^{-1}v_\vartheta^{-1}$.
Here, $v_{\rm{rel}}$ is the relative speed of the orbiter and the disc
matter, while $\Sigma_\ast$ is surface density ascribed to the orbiter
and defined by $\Sigma_\ast=M_\ast/\left(\pi{R^2_\ast}\right)$
(quantities denoted by asterisk refer to the orbiter). At every
collision, a tiny kick changes velocity according to Eq.~(\ref{eq:v'})
and leads to corresponding changes of orbiter's specific (per unit mass)
binding energy $E$, specific angular momentum $L$, its projection $L_z$
onto rotation axis, and radial velocity component $v_r$:
\begin{eqnarray}
 \delta{E} = -{\textstyle{\frac{1}{2}}}\,\delta{v^2} & = &
 {A\over ay} \left(2z -y + xz^{3/2}\right),
 \label{eq:delta1}\\
 \delta{L} = r\,\delta v_{\rm{}t} & = & A \sqrt{ ay \over z}
 \left(x-\sqrt{z}\right), \label{eq:delta2}\\
 \delta{L_z} = r\,\delta v_\varphi & = & A
 \sqrt{ ay \over z}\left( 1 - x\sqrt{z} \right),
 \label{eq:delta3}\\
 \delta{v_r} & = & A \sqrt{ 2z - z^2 -y \over ay};
 \label{eq:delta4}
\end{eqnarray}
here we introduced variables $x=\cos{i}$, $y=1-e^2$,
$z=1+e\cos\omega$, and we expanded right-hand sides of Eqs.\
(\ref{eq:delta1})--(\ref{eq:delta4}) to the first order in $A$ (since we
deal with a small perturbation of the orbit). The collisions occur, in
general, twice per each revolution of the orbiter, at the ascending and
descending nodes, which will be denoted by subscripts ``1'' and ``2'',
respectively. Hydrodynamical drag acts when the orbiter passes across
the nodes. By adding the two contributions one obtains
\begin{eqnarray}
 \rd a &=& 2A_1 a y^{-1} \left(y - 2z + xz^{3/2} \right)
  \nonumber \\
 & & + 2A_2 a y^{-1} \left(y - 4 + 2z + x(2-z)^{3/2}
  \right), \label{eq:dif2a} \\ \rd x &=& A_1 {1 - x^2 \over \sqrt{z}} +
  A_2 {1 - x^2 \over \sqrt{2-z}}\,, \label{eq:dif2x} \\
  \rd y &=& 2A_1
  \left(2z - 2y + {xy \over \sqrt{z}} - xz^{3/2} \right) \nonumber \\
 & & + 2A_2 \left( 4-2z- 2y + {xy \over \sqrt{2-z}} - x(2-z)^{3/2} \right),
  \label{eq:dif2y} \\
 \rd z &=& 2A_1 \left(x \sqrt{z} - z \right) + 2A_2
  \left(2-z - x \sqrt{2-z}\right).
  \label{eq:dif2z}
\end{eqnarray}
Eqs.~(\ref{eq:dif2a})--(\ref{eq:dif2z}) determine the orbital elements
in a parametric form, and we can adopt $x$ as a convenient parameter.
Notice that the time dependence has not yet been involved explicitly.
Temporal evolution can be examined by introducing the orbital period
${\rd}t=2{\pi}a^{3/2}/\sqrt{GM}$.

In further equations we take advantage of the fact that the assumed 
density profile, $\Sigma_\rd=K\left(r/r_{\rm{}g}\right)^s\Sigma_\ast$,
enables us relate $A$'s at two different radii:
\begin{eqnarray}
A_1 &=& K {\left(ay\right)^s \over z^{s+1} }
 \,\sqrt{ 3z - y - 2xz^{3/2} \over 1 - x^2}\,, \\
A_2 &=& K {\left(ay\right)^s \over \left(2 - z\right)^{s+1} }
 \,\sqrt{ 6 - 3z - y -2x (2-z)^{3/2} \over 1 - x^2}\,.
\label{eq:A_1}
\end{eqnarray}

Initially, the orbit is energetically bound or parabolic ($E\geq0$) with
$a=a_0$, $i=i_0$, and $\omega=\omega_0$. Subsequent evolution decreases
inclination and eccentricity, and it causes slow precession in $\omega$.
Naturally, there is a lower limit on inclination, which determines a
moment when integration of our evolutionary equations must be terminated
for at least two reasons. First, the nature of the problem becomes
different once the orbiter is fully embedded in the disc (Lin et al.\
1999; Goldreich \& Tremaine 1980). Second, the assumption (\ref{eq:v'})
about impulsive collisions requires the relative speed to
be supersonic.

Regarding the latter condition, we can write,
for a satellite on a quasi-circular Keplerian orbit,
$v_\rK\approx2.1\times10^{10}
\left(r/r_{\rm{}g}\right)^{-1/2}{\rm{cm\;s}}^{-1}$. Therefore,
$v_{\rm{rel}}\approx4.2\times10^{10}\sin(i/2)
\left(r/r_{\rm{}g}\right)^{-1/2}{\rm{cm\;s}}^{-1}$, which is to be
compared with the speed of sound $v_{\rm{s}}$. For example, in the
standard Shakura-Sunyaev accretion disc (Frank et al.\ 1992) with the
central mass $M=10^8M_\odot$, the accretion rate
$\dot{M}=0.01\dot{M}_{\rm{Edd}}$ ($\dot{M}_{\rm{Edd}}$ is the accretion
rate corresponding to Eddington luminosity), and viscosity parameter
$\alpha=0.1$, one obtains $v_{\rm{s}}\approx
1.27\times10^7\left(r/r_{\rm{}g}\right)^{3/8}{\rm{cm\,s}}^{-1}$, and the
condition $v_{\rm rel}{\gg}v_{\rm{s}}$ then reads
$\sin(i/2)\gg3.1\times10^{-4}\left(r/r_{\rm{}g}\right)^{7/8}$.

Inclined orbits can be further distinguished into two sub-cases:
Those with (i)~a single intersection with the disc plane per revolution,
and (ii)~two intersections per revolution.

\subsection{The initial phase: single intersection}
Before describing the results of numerical integration we recall that
the parametric analytical solution was given in VK under the assumption
that collisions occur only once per revolution, at the inner orbital
node (while the more remote node was assumed to lie beyond the outer
edge of the disc). This situation corresponds to the initial stage with
large eccentricity and appropriate orientation of the ellipse in terms
of $\omega$. In this case, one can verify that Eqs.\
(\ref{eq:dif2a})--(\ref{eq:dif2z}) with $A_2=0$ are equivalent to Eqs.\
(20)--(23) of VK. The corresponding solution reads
\begin{eqnarray}
a & = & zy^{-1}r_0,
 \label{eq:avk} \\
y & = & \frac{B_2\left(1-x^2\right)+x\left(x+4B_1\sqrt{1-x^2}\right)}
 {\left(x+B_1\sqrt{1-x_{0}^2}\right)^4},
 \label{eq:yvk} \\
z & = & \left(B_1\sqrt{1-x^2}+x\right)^{-2},
 \label{eq:zvk}
\end{eqnarray}
where $r_0$, $B_1$, $B_2$ are constants of integration. We recall that,
in Eqs.\
(\ref{eq:avk})--(\ref{eq:zvk}), $x$ stands as parameter. Time does not
enter in mutual relations among $a(x)$, $y(x)$, and $z(x)$.
In accordance with the previous definition, $x$ has a physical meaning
of cosine of inclination, and one can find its temporal dependence by
solving implicit relation
\begin{eqnarray}
t &=& {1 \over K} \int_{x_0}^{x}z^{3/2}(\bar{x})
\left[ {z(\bar{x}) \over a(\bar{x})
y(\bar{x}) } \right]^{s} \left( 1 - \bar{x}^2 \right)^{-1/2}
\nonumber \\
 & & \times \left[ 3z(\bar{x}) - y(\bar{x}) -2\bar{x}
z^{3/2}(\bar{x}) \right]^{-1/2}
\,\rd \bar{x} .
\label{eq:tvk} 
\end{eqnarray}

\subsection{The subsequent phase: two intersections}
The orbit eventually develops two intersections per revolution, and one
has to resort to a numerical treatment of the problem. However, we
checked (see below) for the wide range of starting parameters that the
results do not differ substantially from the case with special
orientation, $\omega_0=\pi/2$ (i.e.\ $z=1$), which can still be treated
analytically. In addition, one can verify that this particular orientation 
of the orbit is linearly stable, or, in other words, the orbital evolution
restores the initial argument of the nodes if slightly perturbed from
$\omega=\pi/2$. This conclusion holds for $\Sigma_\rd$ decreasing with
radius ($s<0$), but it requires a further remark: It was shown
(Vokrouhlick\'y \& Karas 1998b) that the claim about the stability is
not true when the disc gravity is taken into account. In that case the
stability could be maintained only if the orbiter--disc interaction is
substantially stronger than it follows from Eq.~(\protect\ref{eq:v'}),
perhaps due to turbulence or different nature of the collisions. In
fact, one should take the disc gravity into account also for
self-consistency reasons (although we ignore it here), because both
gravitational effects of the disc and its dissipative influence on the
passing orbiter arise from the same distribution of matter.

The solution of the $\omega=\pi/2$ case is
\begin{eqnarray}
a &=& \frac{C_1(1+x)}{\left(1+x\right)^3-C_2\left(1-x\right)},
 \label{eq:az1} \\
y &=& 1-C_2\frac{(1-x)}{\left(1+x\right)^3},
 \label{eq:yz1} \\
z &=& 1,
 \label{eq:zz1} \\
t &=& {1 \over K} \int_{x_0}^{x} {\left[ a(\bar{x}) y(\bar{x})
\right]^{-s}\rd \bar{x} \over \sqrt{ \left( 3 - y(\bar{x}) - 2\bar{x}
\right) \left( 1 - \bar{x}^2 \right) }},
 \label{eq:tz1} 
\end{eqnarray}
where constants $C_1$ and $C_2$ are to be determined from initial values
of $x=x_0$ and $y=y_0$.

\begin{figure}[t]
 \epsfxsize=\hsize
 \centering
 \mbox{\epsfbox{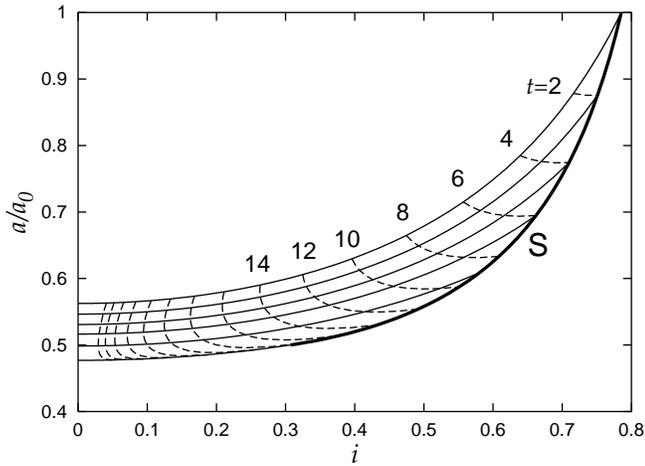}}
 \caption{Semimajor axis $a$ (ordinate, units of $a_0$) versus
 inclination $i$ (abscissa). Inclination decreases from its initial
 value $i_0=\pi/4$ down to zero, while $a$ decreases from $a=a_0$ to the
 final $a=a_{\rm{f}}$ which depends on position of the outer edge of the
 disc. First, the evolution follows the thick solid line {\sf{S}} 
 (single intersection per each orbital revolution), then one of thin 
 solid lines (two intersections per revolution). Dashed curves 
 indicate lines of constant time (arbitrary units). Initial 
 $\omega_0=0$, $e_0=0.5$. See the text for details.
 \label{fig1}}
\end{figure}

\begin{figure}
 \epsfxsize=\hsize
 \centering
 \mbox{\epsfbox{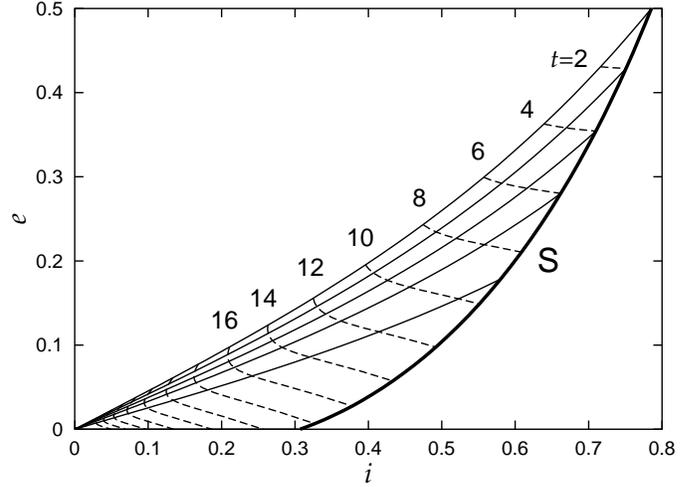}}
 \caption{Eccentricity $e$ versus inclination $i$ for the same initial
 parameters as in Fig.\ \protect\ref{fig1}. Here, $i_0=\pi/4$,
 $\omega_0=0$. Again, the thick solid line shows the initial phase of
 single intersection, while thin lines correspond to the subsequent
 period with two intersections.
 \label{fig2}}
\end{figure}

\begin{figure}
 \epsfxsize=\hsize
 \centering
 \mbox{\epsfbox{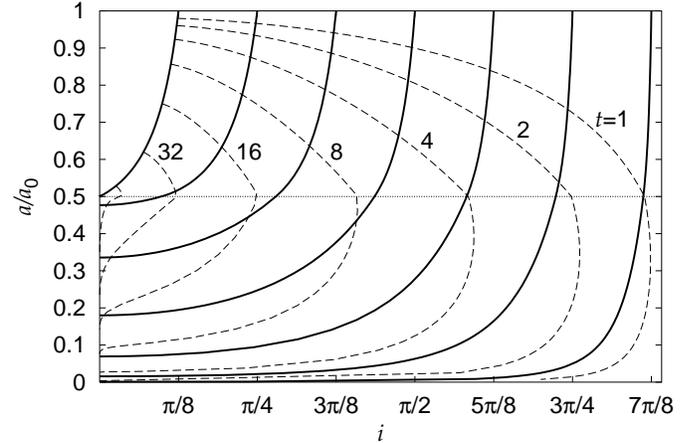}}
 \caption{Semimajor axis $a$ versus inclination $i$ (solid lines) for
 orbits with different $i_0$ (initial inclination). Contours of constant
 time are also shown (dashed; time is given with each curve in arbitrary
 units). Other parameters as in Fig.\
 \protect\ref{fig1}. Trajectories in this figure start with $a=a_0$,
 $i=i_0$, and circularize at $a=\frac{1}{2}a_0$. Then they proceed,
 already with $e=0$, down to terminal radius $a_{\rm{f}}$ at $i=0$.
 \label{fig3}}
\end{figure}

For $e=0$ one obtains $C_2=0$, and Eqs.~(\ref{eq:az1})--(\ref{eq:zz1}) thus
reproduce Rauch's (1995) formula (7) for the final value of semimajor
axis of the circular orbit as function of its initial value and initial
inclination. In the present notation,
\beq
a=C_1\left(1+x\right)^{-2}.
\label{eq:ar}
\eeq
Denoting the initial semimajor axis by $a_{\rm{0}}=a(i_0)$, one
obtains for the constant
\beq C_1=a_{\rm{0}}\left(1+\cos{i_0}\right)^2 =
4\,a_{\rm{0}}\cos^4\left(i_0/2\right).
\eeq

Temporal evolution (\ref{eq:tz1}) can be integrated after specifying
the surface density profile by parameter $s$:
\beq
t = {\pi} K^{-1} C_{1}^{3/2-s}I_{(7-4s)/2},
\label{eq:tz2}
\eeq
where
$I_\gamma(x)=\int_{x_0}^{x}(1+\bar{x})^{-\gamma}\sqrt{1-\bar{x}}\,\rd\bar{x}$.
Integration can be carried out in terms of elementary functions for
suitable values of $s$ (in particular for $s=-3/4$) by applying
recurrent formula
$2(\gamma-1)I_\gamma=(\gamma-3)I_{\gamma-1}-(1+x)^{1-\gamma}\sqrt{1-x}$
and $I_{1/2}=\arcsin{x}$. For other values of $s$, numerical integration
is straightforward.

\section{Results}

We solved Eqs.\ (\ref{eq:dif2a})--(\ref{eq:dif2z}) with different radial
profiles of the surface density and different rotation laws of the disc
matter. Numerical analysis provides us with estimates of accuracy of
the previous results, which were restricted either by assuming low
eccentricity (the final evolutionary stage of the orbit), or, in
turn, by ignoring collisions at the orbital node which is
farther away from the centre (the initial stage). In other words, we
relaxed some assumptions about eccentricity and orientation of the
orbiter's trajectory.

Fig.~\ref{fig1} shows semimajor axis as a function of inclination,
$a(i)$. The orbital evolution proceeds from the initial state,
$a_0=a(i_0)$, to the final $a_{\rm{f}}$ at $i=0$ (when orbiter's orbit
plunges into the disc plane). Thick solid line (denoted by ``{\sf{S}}'')
corresponds to the evolutionary stage with the single intersection,
which is governed by Eqs.\ (\ref{eq:avk})--(\ref{eq:zvk}). (We checked
that this analytical solution, valid during the phase of single
intersections, is reproduced also by the numerical solution which will
be used further.) The orbital evolution follows the line {\sf{S}} to the
point when two intersections occur. The actual moment when this happens
is determined by initial orientation of the ellipse and by the radial
extent of the disc. Subsequent evolution is disjointed from the thick
line, and it goes along one of thin solid lines, which were constructed
according to Eqs.\ (\ref{eq:az1})--(\ref{eq:zz1}). Only this latter part
of the solution depends on the value of $s$ (here we set $s=-3/4$,
corresponding to the outer region in the standard thin disc model).
Dashed lines correspond to $t={\rm{const}}$ in Eq.~(\ref{eq:tz1}). They
are plotted with constant time steps of $\Delta{t}=2$ (arbitrary units)
as indicated in the graph. For clarity of the graph, we do not show the
lines of constant time near $i=0$ because evolution of the system slows
down visibly when inclination is small. Notice that the final radius of
inclined orbits ranges between $0.48\leq{a_{\rm{f}}}/a_0\leq0.57$, lower
values corresponding to the case when the branches with two
intersections develop late in the course of evolution. One can conclude
that even if the moment when two intersections per revolution occur was
unknown (e.g.\ because location of the outer disc edge is left
unspecified), still the final radius could be estimated with the
relative error less than about 0.15. In this plot we choose
$\omega_0=0$, because such initial orientation of the ellipse captures a
long period of evolution when the ellipse has a single intersection per
revolution. We recall that for $\omega=0$ the intersection is at the
pericentre, while the other passage across the disc plane occurs at the
apocentre; cf.\ Fig.~\ref{fig0}.

Fig.~\ref{fig2} is constructed in analogous way as Fig.\ \ref{fig1},
showing eccentricity $e$ as function of $i$. All the curves of $e(i)$
eventually meet together at the origin, which corresponds to circular
orbits of zero inclination (but different $a_{\rm{f}}$'s). We
constructed similar graphs as those in Figs.\ \ref{fig1}--\ref{fig2} but
for different starting parameters, including the counter-rotating
orbits; the results of such analysis are qualitatively the same and can
be quantified by $a_{\rm{f}}$. Complementary to Fig.\ \ref{fig1},
Fig.~\ref{fig3} shows $a(i)$ for different $i_0$. Again, for the sake of
definiteness, the initial ellipse has $e_0=0.5$, $\omega_0=0$, and the
initial pericentre is identified with the outer edge of the disc (i.e.\
radius above which the disc density is considered as negligible). Such
orientation of the orbit could be called as fiducial in our examples
concerning with the capture of a satellite from a remote eccentric
trajectory, initially extending well above $r_{\rm{}h}$. All the orbits
in this example have a single intersection, which is located exactly at
the outer disc edge until $a$ decreases down to half of its initial
value, $a=\frac{1}{2}a_0$ (denoted by the horizontal dotted line). At
this point the second intersection occurs for the first time.

\begin{figure}[t]
 \epsfxsize=\hsize
 \centering
 \mbox{\epsfbox{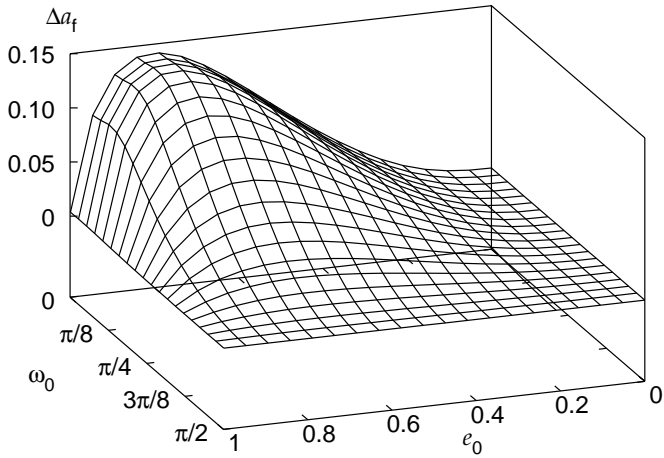}}
 \caption{Relative difference ${\Delta}a_{\rm{f}}(e_0,\omega_0)$ of
 final radius, as defined in Eq.~(\protect\ref{eq:daf}). This figure
 illustrates the fact that Eqs.\
 (\protect\ref{eq:az1})--(\protect\ref{eq:tz1}) estimate the correct
 value of $a_{\rm{f}}$ with accuracy better than 15\% in the whole range
 of eccentricity. 
 \label{fig4}}
\end{figure}

Terminal radius $a_{\rm{f}}$ of a general orbit, with either one or two
intersections per revolution, follows from the numerical solution, as it
was described in previous paragraphs. Eqs.\
(\ref{eq:az1})--(\ref{eq:yz1}) represent a good approximation to the
numerical solution of Eqs.\ (\ref{eq:avk})--(\ref{eq:zvk}), which
becomes exact for $\omega_0=\pi/2$. With this orientation of the
trajectory we recover the above-quoted formula, originally derived for
moderate eccentricity (Rauch 1995): $a_{\rm{f}} {\approx} \frac{1}{4}a_0
\left(1-e_{0}^2\right) \left(1+{\cos}i_{0}\right)^2$. A simple formula
has obvious practical advantages, and one thus wants to quantify its
accuracy. We define the relative error as
\beq
\!{\Delta}a_{\rm{f}}(e_0,i_0,\omega_0)\equiv\frac{a_{\rm{f}} (a_0, e_0,
i_0, \omega_0)-a_{\rm{f}}(a_0,e_0,i_0,\pi/2)}{a_0 } 
\label{eq:daf}
\eeq
and show the surface plot of ${\Delta}a_{\rm{f}}$ in Fig.~\ref{fig4}
for $i_0=5\degr$, $s=-3/4$, and different starting $\omega_0$, $e_0$. It
is evident from the plot that the difference is substantially
reduced when $\omega_0\rightarrow\pi/2$ and/or $e_0\rightarrow0$.
Although this plot was constructed for the particular value of $i_0$, it
turns out that the difference is also reduced for higher initial
inclination.

\begin{figure}
 \epsfxsize=\hsize
 \centering
 \mbox{\epsfbox{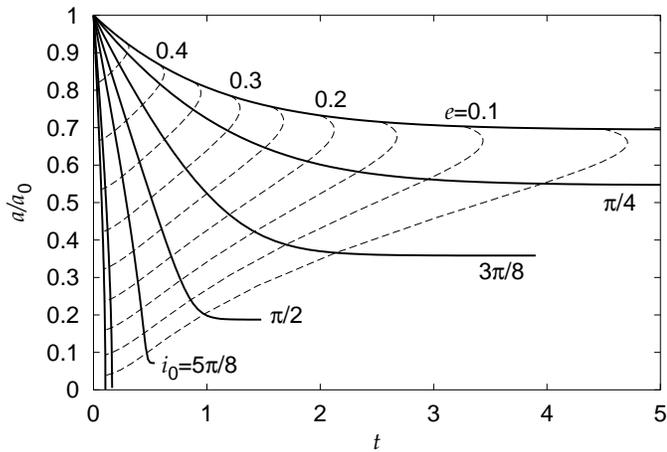}}
 \caption{Time evolution of semimajor axis $a(t)/a_0$ (solid lines).
 Initial inclination is given with each curve. Isolines of constant
 eccentricity (dashed) indicate how the pace of orbital evolution
 eventually slows down in the case of orbits which are almost circular
 and have small inclination. Again, time is in arbitrary units which
 can be scaled with the disc surface density (see the text); $e_0=0.5$,
 $\omega_0=\pi/2$.
 \label{fig5}}
\end{figure}

\begin{figure}
 \epsfxsize=\hsize
 \centering
 \mbox{\epsfbox{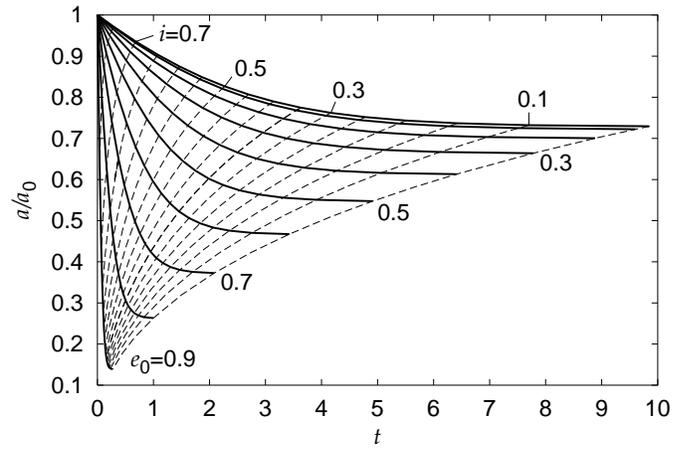}}
 \caption{Time evolution of semimajor axis as in Fig.\
 \protect\ref{fig5} but for $i_0=\pi/4$ and different initial
 eccentricities (solid lines; $e_0$ is given with each curve).
 Isolines of constant inclination are shown 
 with constant step 0.05 (dashed).
 \label{fig6}}
\end{figure}

Our discussion is completed by calculating temporal dependences. Time
evolution of semimajor axis is shown in Figs.~\ref{fig5}--\ref{fig6}
for the disc with $\Sigma_{\rm{d}}\,\propto\,r^{-3/4}$. Here we choose
$\omega_0=\pi/2$, so that the orbit has two branches from the very
beginning. Temporal dependences can be computed by direct evaluation of
Eq.~(\protect\ref{eq:tz1}), or Eq.~(\protect\ref{eq:tz2}) when it
applies, but purely numerical integration of original Eqs.\
(\protect\ref{eq:dif2a})--(\protect\ref{eq:dif2z}) gives the same
results. Characteristic time-scale $\tau$ (in terms of
orbital periods) is determined by ratio of surface densities,
$\tau=\Sigma_{\ast}/\Sigma_{\rm{d}}$. When scaled to the form appropriate for
the gas-dominated region of the Shakura-Sunyaev disc around a
supermassive compact core (Frank et al.\ 1992),
\begin{eqnarray}
\Sigma_{\rm{d}}&=&10^{-4}\,\left(\frac{\alpha}{0.1}\right)^{-4/5}
 \left(\frac{{\protect\dot{M}}}{M_{\odot}/{\rm{yr}}}\right)^{7/10}
 \nonumber \\
 & & \times\,
 \left(\frac{M}{10^8M_{\odot}}\right)^{-1/2}
 \left(\frac{r}{r_{\rm{}g}}\right)^{-3/4}
 \left[\frac{M_{\odot}}{r_{\odot}^2}\right].
 \label{sd}
\end{eqnarray}
From here, the fiducial value of $\tau$ corresponding to a solar-size
satellite ($\Sigma_{\ast}{\approx}M_{\odot}/r_{\odot}^2$) passing across the
inner part of disc (where density has maximum) is of the order $10^4$
revolutions (inverse of the numerical factor in front of rhs.\ of
Eq.~(\ref{sd})). Scaling with the model parameters is also evident:
e.g.\ for an orbiter at $r\approx10^4r_{\rm{}g}$, and the disc with
viscosity parameter $\alpha\approx10^{-3}$, accretion rate
$\dot{M}\approx1{M}_{\odot}$ per year, $\Sigma_{\ast}=\Sigma_\odot$, one
obtains $\tau\approx10^{6}$ revolutions, corresponding to
$\approx10^8\,$yrs. Although the precise time-scale can be computed in a
straightforward manner from Eq.~(\protect\ref{eq:tz1}) above, assuming
the disc model, one should be cautious about interpretation of time
intervals in physical units. In a realistic model, major uncertainty
affecting the constant of proportionality $K$ is connected with poorly
known effects of turbulence (Zurek et al.\ 1996), which tend to enhance
the drag on the orbiter. We recall that $K$ stands only in front of the
time integral, and it determines temporal dependences in absolute units,
but it does not appear in mutual relations among orbital parameters.
Knowledge of $K$ is thus unnecessary for determination of the shape of
the orbit and its evolution in arbitrary units of time, provided that
collisions can be treated as small perturbation.

In the above-given discussion we restricted ourselves to Keplerian
discs. This turns out to be rather important assumption because
non-Keplerian discs can transfer energy to the orbiter. In such a case,
$a(t)$ passes through a minimum at some point and starts increasing
again. For example, such behaviour occurs in a disc with constant
angular momentum density $l$ (Karas \& \v{S}ubr 1999), if one can still
retain other assumptions of our model, namely the impulsive
approximation for collisions ($l={\rm{const}}$ appears in discussion of
discs which acquire substantial geometrical thickness, so that
vertically integrated quantities may be too inaccurate).

\section{Conclusions}
There are two areas where the models, like the one which we discussed
here, are relevant. First, the calculations of the stellar distribution
in galactic cores (Bahcall \& Wolf 1976; Quinlan et al.\ 1995) predict
the presence of density cusps and corresponding integrated-light
profiles around the massive black hole. Stellar orbits and,
consequently, the overall form of the cusps will be modified by the
presence of gaseous environment. We demonstrated that evolutionary
time-scales are typically 1--2 orders of magnitude less than the stellar
life time, and the influence of the disc should be thus accounted. This
is true especially for active galactic nuclei with enhanced accretion
rates, while in the center of Milky Way the gaseous environment is too
rarefied to exert any substantial hydrodynamical drag on stars. The
smallest resolved scales ($\approx0.1^{\prime\prime}$) do not allow
direct observational comparisons at present, but this may be possible
when better resolution is available (although the direct tracking of
individual stars does not seem possible in near future; Kormendy \&
Richstone 1995).

Another possible application concerns the predicted gravitational
waveform patterns from inspiralling satellites. It was suggested
(Chakrabarti 1996) that the wave emission may be significantly altered
if the satellite moves inside a super-Eddington
($\dot{M}\approx10^3\dot{M}_{\rm{Edd}}$) flow and passes through shocks
there. This could be an interesting possibility but it represents rather
exceptional situation. When most of the motion takes place outside the
disc, as in our case, the time-scales of the orbital evolution due to
the presence of the disc are always much longer than the orbital period,
even for quite high accretion rates
($\dot{M}\approx\dot{M}_{\rm{Edd}}$), and the production of
gravitational waves is virtually unaffected. One should however notice
that gaseous environment helps to capture the satellite and bring it
down to the low orbit with small inclination, the situation which is
needed for the efficient wave emission.

We solved Eqs.\ (\ref{eq:dif2a})--(\ref{eq:dif2z}), which determine the
orbital evolution of a body on an eccentric trajectory evolving
adiabatically due to dissipative collisions with the disc. In the
general case it is quite obvious from the form of these equations that
one has to resort to the numerical approach, but we checked with discs
obeying reasonable power-law surface density profiles
($-1\leq{s}\leq-0.5$) and Keplerian rotation law that evolution of the
true trajectory is well approximated by the corresponding orbit with
$\omega_0=\pi/2$. This case can be integrated analytically [Eqs.\
(\ref{eq:az1})--(\ref{eq:tz1})]. Approximate terminal radii of
circularized orbits agree with those obtained by numerical integration
within precision better than 15\%. This conclusion can be put in other
words, namely, effects of the disc self-gravity, of different rotation
laws, and of mutual collisions among orbiters seem to influence the
evolution more substantially than do exact properties of the disc
material.

Compared with previous papers, our present discussion has been more
general on the level of semi-analytical treatment of the orbits with
arbitrary orientation, for which either one or two intersections develop
in the course of evolution. In particular we examined how the power-law
index of the disc surface-density profile enters in equations for the
orbital shape and time evolution of the parameters. It should be quite
obvious that we picked up one aspect of a complex problem of the
long-term orbital evolution, which will be build into a mosaic to form
the complete picture in astrophysically more realistic models. Various
compelling processes were already considered separately. Now we need to
take them jointly into account, especially the effects of resonances
between the satellite and the disc matter, corrections due to general
relativity (of the gravitational field of the central object and energy
losses of the orbiter), tidal effects, and gravitational interaction
with other satellites. This will certainly require mostly numerical
approaches for which analytical estimates of individual processes and
toy models, like the present one, provide useful guidelines.

\begin{acknowledgements}
\def\lb#1{{\protect\linebreak[#1]}}
We are grateful to the referee for suggestions which helped
us to improve the original version of our paper. We also thank
D.~Vokrouhlick\'y for clarifying discussions and we acknowledge
support from the grants GACR 205/\lb{2}97/\lb{2}1165 and
202/\lb{2}99/\lb{2}0261. V.\,K. thanks for hospitality of the
International School for Advanced Studies in Trieste.
\end{acknowledgements}
  

\end{document}